\documentclass[conference,a4paper]{IEEEtran}
\usepackage[utf8]{inputenc}
\usepackage[english]{babel}

\usepackage{cite}
\usepackage{amsmath,amssymb,amsfonts}

\usepackage[ruled, lined, linesnumbered, commentsnumbered, longend]{algorithm2e}
\usepackage{algpseudocode}

\usepackage{graphicx}
\usepackage{textcomp}
\usepackage{xcolor}
\usepackage{soul}
\usepackage{enumerate}
\usepackage{comment}

\usepackage{multirow}
\usepackage{tabu}
\usepackage{float}
\usepackage[inline]{enumitem}
\usepackage{amsmath}
\usepackage{mathtools}
\usepackage{booktabs}

\usepackage[autostyle]{csquotes}
\usepackage[utf8]{inputenc}

\usepackage{textcase}

\usepackage[left=1.62cm,right=1.62cm,top=1.9cm]{geometry}

\definecolor{boristext}{rgb}{0.3, 0.36, 0.88}
\definecolor{boriscomments}{rgb}{0.83, 0.0, 0.0}
\definecolor{davidcomments}{rgb}{0.0, 0.0, 0.83}

\usepackage{bbm}

\begin{document}

\bstctlcite{IEEEexample:BSTcontrol}

\title{Improving Wi-Fi 8 Latency with\\Coordinated Spatial Reuse}

\author{
\IEEEauthorblockN{David Nunez$^{\flat}$, Francesc Wilhelmi$^{\flat}$, Lorenzo Galati-Giordano$^{\star}$, Giovanni Geraci$^{\sharp}$\,$^{\flat}$, and Boris Bellalta$^{\flat}$\vspace{0.1cm}}
\IEEEauthorblockA{$^{\flat}$\emph{Department of Engineering, Universitat Pompeu Fabra, Barcelona, Spain}}
\IEEEauthorblockA{$^{\star}$\emph{Radio Systems Research, Nokia Bell Labs, Stuttgart, Germany}}
\IEEEauthorblockA{$^{\sharp}$\emph{Telef\'{o}nica Scientific Research, Barcelona, Spain}}
\IEEEauthorblockN{\thanks{Corresponding author: \emph{david.nunez@upf.edu}.}
}}
\maketitle


\begin{abstract}
IEEE 802.11 networks continuously adapt to meet the stringent requirements of emerging applications like cloud gaming, eXtended Reality~(XR), and video streaming services, which require high throughput, low latency, and high reliability. To address these challenges, Coordinated Spatial Reuse (Co-SR) can potentially contribute to optimizing spectrum resource utilization. This mechanism is expected to enable simultaneous transmissions, thereby boosting spectral efficiency in dense environments and increasing the overall network performance. In this paper, we shed light on the performance of Co-SR for Wi-Fi~8 networks. For that, we propose an implementation of Co-SR aligned with ongoing Wi-Fi 8 standardization efforts. The evaluation is done on a Wi-Fi simulator, which allows us to study the performance of the proposed Co-SR mechanisms in relevant scenarios. The results obtained in a Wireless Local Area Network (WLAN) consisting of four APs show delay reduction with Co-SR ranging from 31\% to 95\% when compared to Distributed Coordination Function (DCF).
\end{abstract}


\section{Introduction}

Emerging applications such as eXtended Reality (XR), high-quality holographic video, and zero-delay file exchange for remote collaboration require wireless networks to extend their capabilities further. These applications demand high throughput, low latency, and highly reliable connections to ensure seamless user experiences \cite{WBA_2025_report}. At the same time, technologies like Wi-Fi continue to play a dominant role in providing connectivity for these use cases, particularly given the increasing reliance on unlicensed spectrum \cite{OugGerPol2024}. Dense Wi-Fi deployments, where multiple Access Points (APs) share the same frequency resources, experience high levels of contention and collisions, leading to degraded throughput, increased latency, and compromised reliability. New techniques are necessary to optimize channel access and resource utilization. 

To address the challenges imposed by next-generation applications in high-density Wireless Local Area Network (WLAN) scenarios, the future IEEE 802.11bn amendment \cite{UHRobjectives} will introduce Multi-Access Point Coordination (MAPC). This framework is envisioned to enhance the overall latency and reliability performance, by alleviating channel access contention. The definition of a MAPC framework started with IEEE 802.11be~\cite{AdrianGarciaSurvey} but was then postponed until IEEE 802.11bn~\cite{UPF_Nokia_wifi8}. Since then, multiple flavors of MAPC have been proposed to address, among others, the coordination protocols and signaling, setup establishment, and transmission management~\cite{MAPC_mentor_MediaTek_1217r2, MAPCframework_mentor_LG_1514r1}.

MAPC enables the introduction of different methods that perform coordinated resource allocation, such as Coordinated Time Division Multiple Access (Co-TDMA), where APs agree to split time resources~\cite{Val_2025}, or Coordinated Restricted Target Wake Time (Co-RTWT) where scheduled, interference-free communication time slots are reserved for devices, improving power efficiency and real-time application performance~\cite{haxhibeqiri2024coordinated}. In certain scenarios, higher performance can also be achieved by enabling simultaneous transmissions through Coordinated Spatial Reuse (Co-SR) or Coordinated Beamforming (Co-BF). Among these techniques, Co-SR stands out as a particularly promising scheme for reducing latency and, in most scenarios, increasing throughput~\cite{mypaper_throughputAnalisis}.

Closer in spirit to this work, \cite{mio_MAPC_groups_scheduling} introduced a framework in which compatible multi-AP groups are scheduled to perform simultaneous transmissions within a TDMA-slotted Transmission Opportunity (TXOP). When it comes to Co-SR, the authors in \cite{TXOPsharingPaper} evaluated the throughput of networks implementing Co-TDMA and Co-TDMA with SR (Co-TDMA/SR), showing that Co-TDMA/SR leads to throughput gains between 50\% and 140\%, depending on the scenario, when compared to Co-TDMA. As for the analytical characterization of Co-SR, a model based on Markov chains is introduced in \cite{francesc_wifi8_markov}, to compute the Co-SR's throughput in comparison to the Distributed Coordination Function (DCF).

Notably, existing literature lacks a comprehensive performance evaluation of MAPC networks, specifically focusing on delay metrics with Co-SR as a key enabler. In \cite{mypaper_throughputAnalisis}, the performance of a MAPC framework utilizing Co-SR is evaluated, providing a detailed analysis and closed-form expressions for throughput computation. This work builds upon that study by enhancing the previously proposed Co-SR group creation algorithm, aiming to find an optimal solution that balances capacity and fairness. Furthermore, it presents a more in-depth evaluation of Co-SR through simulations, which allows assessing the delay gains achieved in comparison with DCF.  

By way of a summary, this work proposes a Co-SR implementation for Wi-Fi 8 networks that leverages spatial reuse groups of AP-Station (STA) pairs. These groups are selected by solving an optimization problem that maximizes the transmission capacity, considering their transmission probability and guaranteeing that all the AP-STA pairs are granted a minimum number of transmission opportunities, so that fairness is preserved. A performance evaluation is carried out for enterprise-like deployments with four APs and randomly distributed stations, allowing us to show that our proposed Co-SR design outperforms the baseline DCF mechanism. We also analyze the impact of limiting the maximum number of simultaneous multi-AP transmissions to two, as currently proposed for IEEE 802.11bn, versus an unconstrained number of coordinating APs, which allows us to explore the potential of Co-SR beyond Wi-Fi 8.


\section{Challenges and Directions of MAPC and Co-SR}

MAPC is intended to improve network performance by enabling collaboration among APs, reducing co-channel interference, and optimizing resource allocation. Unlike conventional Wi-Fi networks, where each AP manages its transmissions independently, MAPC introduces mechanisms that allow APs to share information, synchronize transmissions, and make joint decisions regarding channel access, power control, or resource allocation. 

For its part, Co-SR optimizes spectrum utilization by enabling concurrent transmissions, which can significantly boost spectral efficiency improving throughput, latency, and reliability. This is particularly beneficial in networks with a high density of APs and STAs, where traditional methods like DCF struggle to manage traffic efficiently, and queued packets wait long periods before being transmitted. 

\subsection{Challenges}

Implementing MAPC techniques presents several challenges, ranging from signaling overhead and scalability to fairness in scheduling and real-time adaptability. Addressing these challenges is crucial for ensuring that coordination mechanisms remain effective and scalable in dense and heterogeneous network environments. Below, we outline some of the main open challenges associated with MAPC and Co-SR, highlighting key issues that need to be addressed for a successful implementation.

\begin{itemize}
    \item \textit{Signaling overheads:} Effective coordination requires control messages for the discovery and setup of the mechanism, plus periodic exchanges of enabling information (e.g., interference levels, transmission schedules, or resource allocation). This can introduce significant signaling overhead, particularly in dense Wi-Fi deployments where multiple APs and STAs operate simultaneously.
    \item \textit{Scalability:} As the number of APs and STAs increases, scheduling and decision-making become more challenging, as the complexity of coordination groups grows in a combinatorial manner. Device heterogeneity exacerbates these issues, as modern networks typically comprise Wi-Fi 4 to Wi-Fi 7 devices with different capabilities.
    \item \textit{Fairness in resource allocation:} In MAPC, where scheduling decisions are explicitly coordinated, prioritizing certain AP-STA links to optimize overall network efficiency may unintentionally starve others of resources. Ensuring a fair balance between efficiency and equitable access remains a significant research challenge.
\end{itemize}
 
When it comes to the specific challenges associated with the implementation of Co-SR, we outline the following ones:
  
\begin{itemize}
    \item \textit{Interference measurement:} Accurate measurement of interference levels is crucial to make informed decisions about spatial reuse opportunities. This requires collecting Received Signal Strength Indicator (RSSI) from neighboring APs and STAs, which can introduce significant overhead, especially in high-density environments. The frequency of these measurements, the type of information collected, and the processing requirements must be carefully balanced to avoid excessive network overhead while maintaining accurate interference estimation. Environmental dynamics further strain system responsiveness, i.e., changing channel conditions render precomputed coordination strategies obsolete within near-second timescales, making real-time measurement and adaptation even more complex.
    \item \textit{Modulation and Coding Scheme (MCS) selection:} The effectiveness of Co-SR depends heavily on selecting near-optimal MCS for transmissions that overlap in the same geographical area, so that the Bit Error Probability (BER) is effectively reduced while preserving high enough data rates. Since MCS selection depends on Signal-to-Interference-plus-Noise Ratio (SINR), designing adaptive algorithms that can dynamically adjust MCS in response to changing network conditions is still an open research problem.
    \item \textit{Scenario dependency:} Co-SR's effectiveness depends on network topology, device density, mobility, and traffic patterns. In dense environments with heavy congestion, aggressive spatial reuse can lead to excessive interference, reducing overall network performance instead of improving it, while in sparsely populated networks, coordination benefits may be minimal, raising questions about under which conditions Co-SR could be applied.
    \item \textit{Coordination scope and decision-making:} Deciding the subset of devices that coordinate and their allocated resources is non-trivial. While offering greater control, centralized approaches may need additional infrastructure to exchange the signaling information and enforce commands, whereas distributed approaches require APs to make decisions while avoiding conflicts independently. Achieving a balance between control and autonomy is essential for effective spatial reuse in large-scale Wi-Fi networks.
\end{itemize}

\subsection{IEEE 802.11bn Directions}

The IEEE 802.11bn Task Group (TGbn) is leading the development of the new amendment. For MAPC, TGbn will define a foundational framework, including frames and basic procedures, while leaving room for manufacturers to implement their solutions. Furthermore, the list of available schemes to be used in the context of MAPC is still under discussion, being so far Co-SR, Co-BF, Co-TDMA, and Co-RTWT the most likely to be included. 

Based on recent contributions and discussions to the TGbn, it is expected that MAPC features will be implemented as follows. APs participating in MAPC can utilize management frames to advertise and discover the capabilities or parameters of specific schemes. Once APs identify potential coordination partners, they can exchange individually addressed management frames to establish agreements and negotiate scheme parameters. For interoperability purposes, 802.11bn should also specify AP-to-AP frame formats, allowing APs to coordinate transmissions either over-the-air or via the Distribution System (DS). 

Regarding Co-SR, it is expected to operate with a maximum of two simultaneous transmissions, leaving a higher number for future amendments. 


\section{Proposed Co-SR Mechanism Model}\label{section:CSR}

Our envisioned Co-SR approach builds upon DCF while incorporating group formation and periodic updates based on exchanged RSSI reports between APs to enhance spatial reuse efficiency. Coordination occurs on a per-TXOP basis, where signaling mechanisms facilitate real-time adjustments. Periodic updates (e.g., every few seconds) to the coordination groups help to adapt to changing channel and traffic conditions, maintaining a balance between control overhead and performance gains.

\subsection{Co-SR Operation}

We propose a Co-SR mechanism that enhances network throughput and latency by enabling concurrent transmissions among groups of coordinated APs. This mechanism, inspired by the ongoing discussions at the TGbn, operates within the DCF framework, facilitating TXOP sharing among compatible APs (see Section~\ref{section:group_formation}). Specifically, when an AP—designated as the \textit{sharing AP}—wins the channel contention (i.e., its backoff counter reaches zero before the one from the other contenders), it initiates a multi-AP transmission. The sharing AP selects one of its associated STAs (according to the scheduling mechanism in Section~\ref{section:group_scheduling}) and triggers a pre-defined group of AP-STA pairs from other Basic Service Sets (BSSs) that are compatible with it for simultaneous transmission. This phase, lasting for a duration of \(T_{\rm MAPC}\), involves channel reservation and the exchange of information about the set of devices and transmission parameters for the upcoming coordinated TXOP. These parameters can include bandwidth, MCS, and TXOP duration. 

\begin{figure}
    \centering
    \includegraphics[scale=0.55]{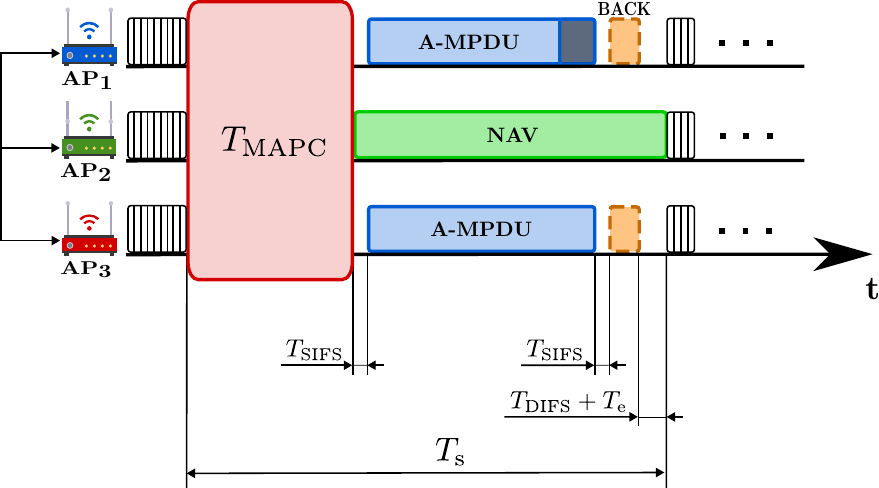}
    \caption{Example of the proposed MAPC operation in a small overlapping BSS deployment where AP$_{1}$ and AP$_{3}$ are selected to participate in a coordinated TXOP, due to their spatial reuse compatibility.}
    \label{Fig:timeline}
\end{figure}

An example of a Co-SR transmission is illustrated in Fig.~\ref{Fig:timeline}. Initially, coordinated APs (AP$_{1}$, AP$_{2}$, and AP$_{3}$) contend for accessing the channel using DCF. Upon winning the TXOP, one of the coordinated APs (e.g., AP$_{1}$) initiates a multi-AP transmission during the \(T_{\rm MAPC}\). From that moment on, a coordinated transmission (e.g., involving AP$_{1}$ and AP$_{3}$) takes place through the exchange of Aggregate MAC Protocol Data Units (A-MPDUs) and Block ACKs (BACKs) between the devices of each involved BSS. Meanwhile, non-transmitting APs (e.g., AP$_{2}$) set their Network Allocation Vector (NAV) and freeze their backoff countdowns.  

\subsection{Multi-AP Group Formation}\label{section:group_formation}

An SR group consists of several APs and STAs pairs that are compatible, i.e., pairs of devices that can transmit simultaneously without causing the SINR at the receiver to fall below a threshold $\gamma_{\textrm{CE}}$ (see Section~\ref{section:performance_eval} for further details). The groups are defined based on the RSSI perceived by each receiver from the neighboring APs.\footnote{Owing to time division duplexing and channel reciprocity, downlink RSSI can be estimated from measurements on uplink frames (data or ACK frames).} These groups are computed by a logical controller, which may be physically located within an AP, an external server, or even in the cloud. The controller is responsible for periodically receiving fresh RSSI data, recalculating the Co-SR groups, and disseminating updates to all APs in the network. In low-mobility Wi-Fi scenarios, RSSI variations occur at a slow timescale, minimizing the overhead associated with data acquisition and transmission to the controller. While the computation of Co-SR groups is centralized, our proposed mechanism ensures that channel access and per-transmission signaling adhere to standard 802.11 operations. This design guarantees fair coexistence with other existing Wi-Fi networks. 

To manage the group formation efficiently, we start by identifying all possible AP-STA (associated) pairs. From there, we explore different ways to group them so they can transmit simultaneously. As mentioned earlier, managing coordination groups becomes increasingly challenging in dense Wi-Fi networks, where the number of coordinated devices can be large. Hence, reducing the number of devices per group can simplify things, making coordination more efficient. While the balance between performance and computational complexity is beyond the scope of this work, we outline two possible approaches for group formation, based on whether or not the number of members per group is limited:

\begin{enumerate}
    \item Unconstrained (UNC): There is no limit on the size of the groups that can perform simultaneous transmissions, allowing as many AP-STA pairs as possible to transmit together.
    \item IEEE 802.11bn Approach (MAX2): Following directions from TGbn discussions, groups for performing simultaneous transmissions are restricted to a maximum of two AP-STA pairs at a time.
\end{enumerate}

\subsection{Multi-AP Transmissions Scheduling}
\label{section:group_scheduling}

Not all the compatible groups of AP-STA pairs lead to the same performance and some allow more efficient data transmission than others. For that reason, to decide on the scheduling of transmission groups, we focus on: \textit{1)} how often each group accesses the channel (because one of its members has won the contention), and \textit{2)} how many packets can be aggregated by the AP-STA pairs in those groups. While selecting a high number of devices in a given simultaneous transmission is preferable, the quality of their individual transmissions depends on the total interference received from the other transmitters within the group, which also affects the chosen MCS. Thus, we define the quality of each group as the product of its transmission probability and the estimated number of packets transmitted within a given TXOP duration by the AP-STA pairs it contains.

The selection of the best groups to be scheduled is then computed by solving the optimization problem that finds those groups that maximize the product of the aforementioned number of transmitted packets considering all the selected groups in the process. Furthermore, to guarantee that all the AP-STA pairs are granted a minimum number of transmission opportunities, we establish a constraint for all the AP-STA pairs to be selected the same number of times. For the sake of simplicity and without loss of generality, we set this value equal to one for the remainder of this paper, which means that, for a given deployment, an AP-STA pair will only appear once in a single group. 

\section{Performance Evaluation}\label{section:performance_eval}

We developed a MATLAB-based simulator designed to evaluate the performance of next-generation Wi-Fi networks, with a particular focus on features of the IEEE 802.11bn amendment.\footnote{This simulator will be openly available in https://github.com/dncuadrado/802.11bn-CoSR-Simulator ---once the paper is accepted--- to researchers as a reliable tool for analyzing MAPC performance in Wi-Fi networks.} 

\subsection{Simulation Scenario and Parameters}

\begin{figure}[t!!!!!!!!!!]
    \centering
    \includegraphics[scale=0.65]{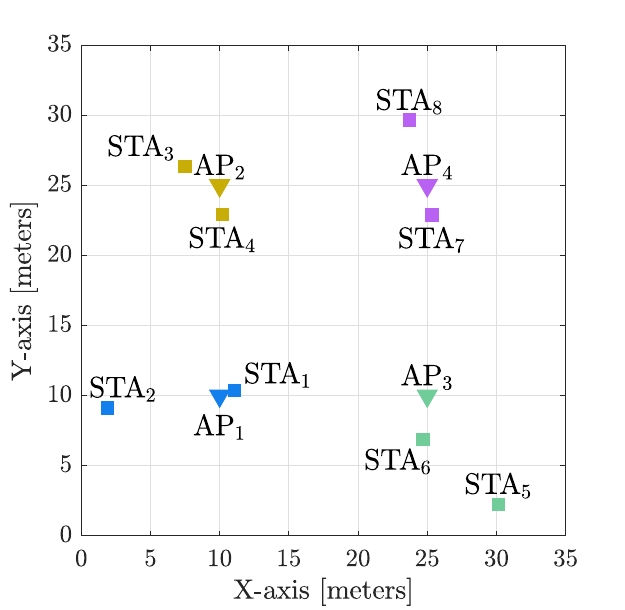}
    \caption{Example Deployment that consists of four APs separated to a distance $d_\text{AP-AP}=15$ meters. Two STAs are associated to each AP.}
    \label{Fig:scenario}
\end{figure}

We consider four overlapping BSSs deployed throughout a squared scenario (see deployment in Fig.~\ref{Fig:scenario}), with an inter-AP distance of $d_\text{AP-AP}$ meters, and where downlink transmissions are held. The APs contend to access the same 80 MHz channel using DCF, so packet collisions may occur during that process. We consider that each AP has two associated STAs. The STAs are randomly placed $d_\text{AP-STA}$ meters away from their corresponding AP.

Each AP manages separate packet queues for its connected STAs. Without loss of generality, we assume each STA handles only a single data stream at a time. We analyze scenarios where the traffic load per STA is 90\% of the throughput achievable by the STA with the lowest MCS for a given scenario when DCF is employed. All queued packets are processed in the order they arrive, following a First-In, First-Out (FIFO) rule. We consider two traffic models:
\begin{enumerate}
    \item \textit{Poisson traffic:} This model represents a continuous flow of packets. 
    The time between packet arrivals varies randomly, following an exponential distribution, meaning some packets may arrive quickly, while others may be spaced further apart.
    \item \textit{Bursty traffic:} Typical internet traffic, like web browsing or cloud services, generates data in bursts rather than a continuous stream. This behavior is well represented by an ON/OFF Markovian model, where data packets are sent only during active (ON) periods, while idle (OFF) periods introduce gaps in transmission. The duration of both ON and OFF periods, defined as $T_{\rm ON}$ and $T_{\rm OFF}$, respectively,  follows an exponential distribution.
\end{enumerate}

The transmission rate achieved in a given AP-STA transmission depends on the selected MCS (we consider 802.11be MCS values), which is computed from the estimated RSSI at the receiver. The path loss is modeled using the IEEE 802.11ax Task Group model for Enterprise Scenarios \cite{pathloss}. Besides, we consider that all the APs use the same fixed transmit power. As a result, stations close (far) from their AP are served using a higher (lower) MCS. Moreover, the maximum number of aggregated packets in each A-MPDU depends on the MCS and the maximum TXOP duration ($T_\text{max})$. For an SR transmission to be successful, the SINR at the receiver must be above a predefined capture effect threshold, $\gamma_{\textrm{CE}}$ \cite{lee2007CaptureEffect}.

Table~\ref{tab:simulation_paramters} collects the simulation parameters used in the evaluation.

\begin{table}
    \caption{Simulation Parameters.}
    \label{tab:simulation_paramters}
    \scriptsize
    \begin{center}
        \begin{tabular}{@{\hspace{0.2cm}}l@{\hspace{0.1cm}}l@{\hspace{0.2cm}}ll}
        \toprule
           \textbf{Parameter}  & \textbf{Description} & \textbf{Value} \\ \midrule
            $K$ & Number of BSSs & 4 \\
            $S_{k}$ & Number of STAs per BSS & 2 \\
            $d_{\textrm{AP-AP}}$ & Inter-AP distance [meters] & \{10, 15, 20\} \\
            $d_{\textrm{AP-STA}}$ & Distance between AP and STAs [meters] & [1, 10] \\
            $\textrm{BW}$ & Bandwidth [MHz] & 80 \\
            $W_{\textrm{n}}$ & Number of  walls (every 10 meters) & \{0, 1, 2\}  \\
            $N_{\textrm{sc}}$ & Number of data subcarriers & 980 \\
            $N_{\textrm{ss}}$ & Number of spatial streams & 2 \\
            $f_{\textrm{c}}$ & Carrier frequency [GHz] & 6 \\
            $T_{\textrm{OFDM}}$ & OFDM symbol duration [$\mu$s] & 12.8 \\
            $T_{\textrm{GI}}$ & Guard interval duration [$\mu$s] & 0.8 \\
            $T_{\textrm{max}}$ & Max. TXOP duration [ms] & 5 \\ 
            $T_{\textrm{MAPC}}$ & Coordination overheads [$\mu$s] & 286\\
            $T_{\textrm{BACK}}$ & Block ACK duration [$\mu$s] & 100 \\
            $T_{\textrm{SIFS}}$ & Duration of a SIFS slot [$\mu$s] & 16 \\
            $T_{\textrm{DIFS}}$ & Duration of a DIFS slot [$\mu$s] & 34 \\
            $T_{\textrm{c}}$  & Duration of a collision slot [$\mu$s] & 137 \\ 
            $T_{\text{e}}$ & Duration of an empty slot [$\mu$s] & 9 \\
            CW$_{\textrm{min}}$  & Min contention window & 15 \\
            CW$_{\textrm{max}}$  & Max contention window & 1023 \\
            $\gamma_{\textrm{CE}}$ & Capture effect threshold [dB] & 15 \\
            $P_{\textrm{max}}$ & Transmission power [mW] & 200 \\
            $W$ & Noise power  [Watts] & $3.2 \times 10^{\textrm{-13}}$  \\
            CCA & Clear Channel Assessment [dBm] & -82 \\
            $T_{\textrm{ON}}$ & Bursty traffic ON period mean duration [ms] & 1 \\
            $T_{\textrm{OFF}}$ & Bursty traffic OFF period mean duration [ms] & 10 \\
            $T_{\textrm{sim}}$ & Simulation duration [s] & 5 \\
            $L$ & Length of single data frame [bits] & $12\times 10^{\textrm{3}}$ \\ \bottomrule
        \end{tabular}
    \end{center}
\end{table}


\subsection{Example Deployment}

\begin{figure}[t!]
    \centering
    \includegraphics[scale=0.44]{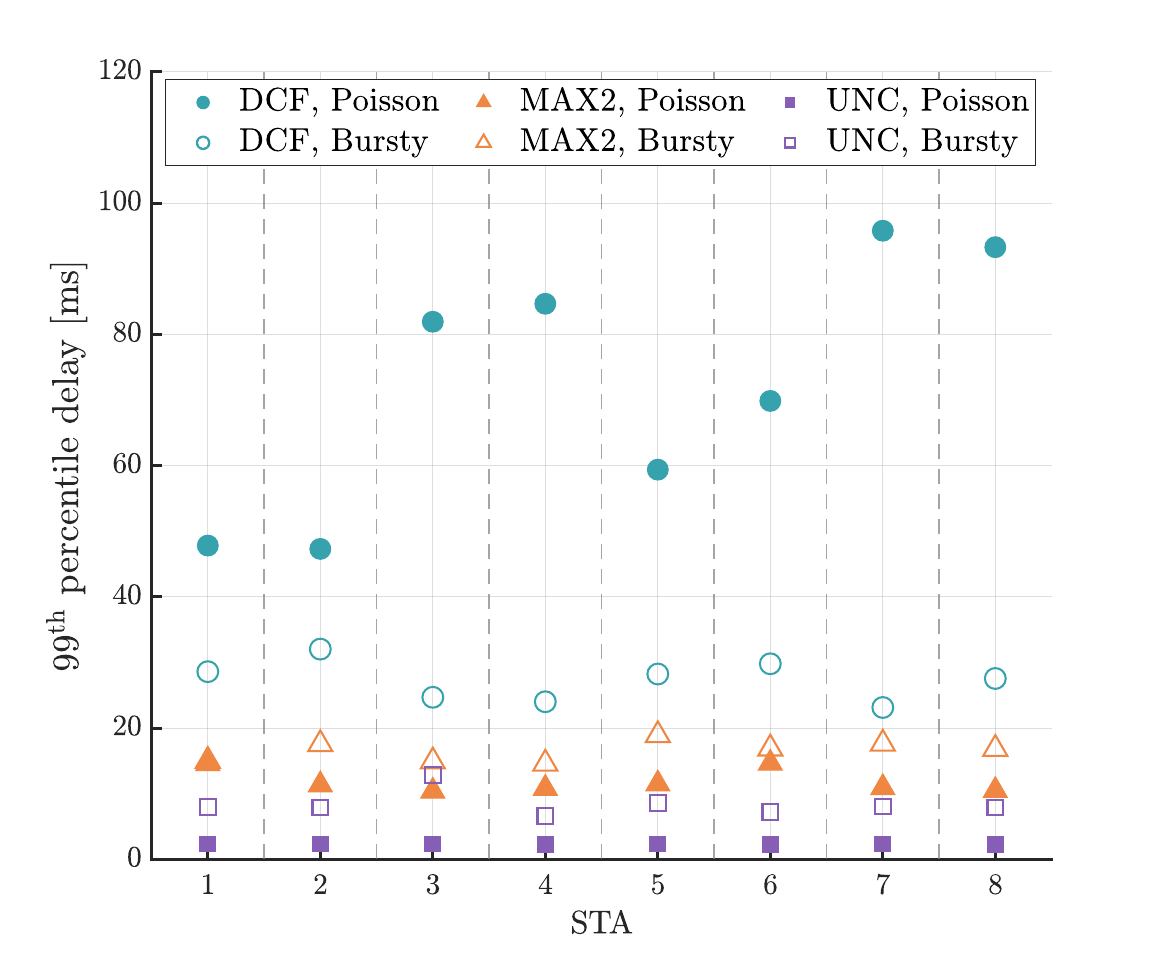}
    \caption{STA delay in the deployment of Figure~\ref{Fig:scenario}, for DCF, MAX2 and UNC with Poisson and Bursty traffic.}
    \label{Fig:STAdelayScenario}
\end{figure}

Let us examine the delay experienced by the STAs in the deployment illustrated in Fig.~\ref{Fig:scenario}, where four APs are positioned, each maintaining a separation of $d_\text{AP-AP}=15$ meters (square side) from the others. This particular distance allows for the generation of appealing Co-SR interactions among BSSs. Fig.~\ref{Fig:STAdelayScenario} presents the $99^{\text{th}}$ percentile delay when using DCF, MAX2, and UNC under both Poisson and Bursty traffic models. 

Notably, DCF is less efficient with Poisson traffic than with Bursty traffic, as it struggles to clear the queues when packets arrive in a continuous stream. In contrast, both MAPC-based approaches significantly reduce the $99^{\text{th}}$ percentile delay compared to DCF, across all STAs and traffic models. For $\textrm{STA}_{1}$, the delay under MAX2 is nearly identical (less than 1 ms difference) for Poisson and Bursty traffic, causing both markers (orange triangles) to overlap. The key advantage of MAPC over DCF stems from Co-SR's ability to enable concurrent transmissions, resulting in more efficient data delivery. Among the MAPC strategies, UNC achieves the best performance, reducing the delay by 95\% compared to DCF, and 78\% compared to MAX2 under Poisson traffic, while achieving reductions of 50\% and 14\% for Bursty traffic, respectively. This significant improvement is enabled by the ease of compatible AP-STA pairs grouping in this Co-SR-friendly deployment. In fact, only two groups are required to serve all eight STAs when no group-size constraints are imposed:
$\{\textrm{STA}_{1}, \textrm{STA}_{3}, \textrm{STA}_{6}, \textrm{STA}_{7}\}$ and $\{\textrm{STA}_{2}, \textrm{STA}_{4}, \textrm{STA}_{5}, \textrm{STA}_{8}\}$,
resulting in an aggregate throughput gain of 284\% over DCF and approximately 100\% compared to MAX2.

 
\subsection{Effect of the Inter-AP Distance}

\begin{figure*}
    \centering
    \includegraphics[scale=0.55]{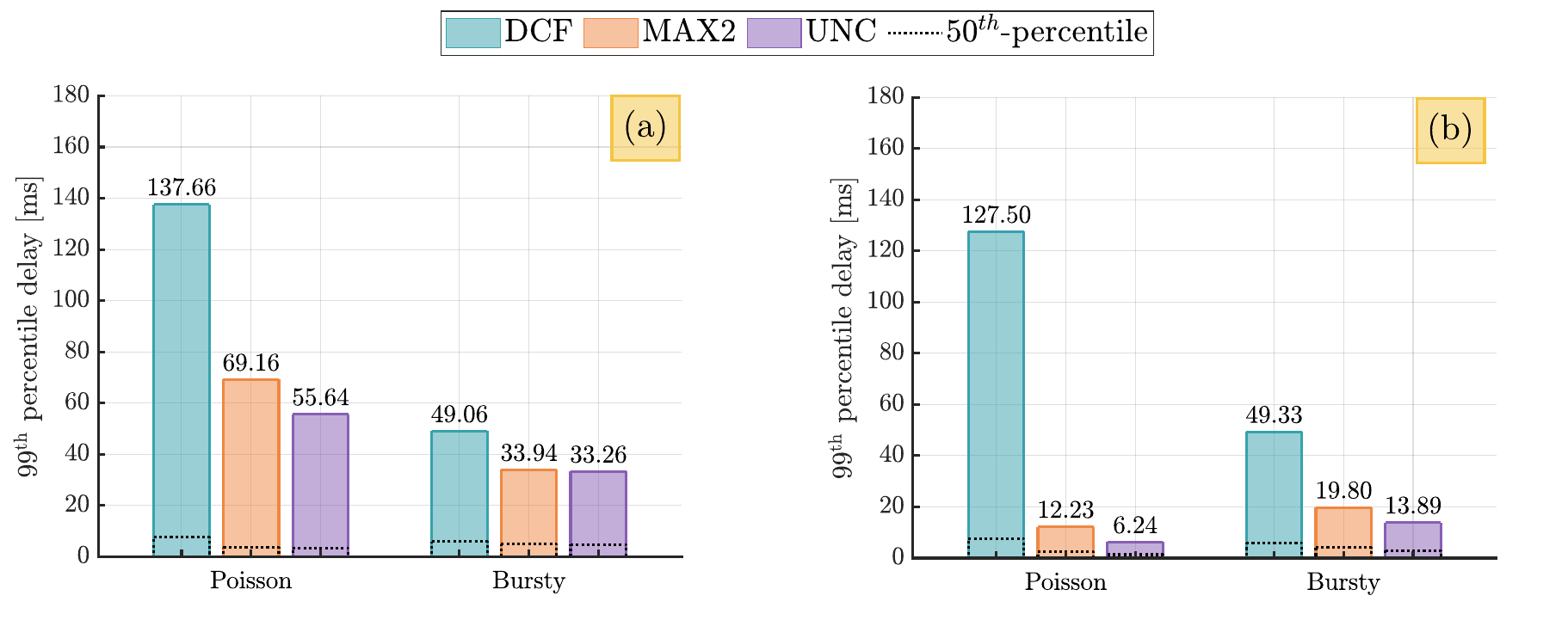}
    \caption{$99^{\text{th}}$ and $50^{\text{th}}$  percentile delay of packets for Poisson and Bursty flows in the 100 analyzed random deployments. Results were obtained for two different distances between APs: (a) $d_\text{AP-AP}=10$ meters and (b) $d_\text{AP-AP}=20$ meters.}
    \label{Fig:randomDeployments}
\end{figure*}

To gain deeper insights into these scenarios, we analyze packet delay across 100 different deployments, varying the inter-AP distance, $d_\text{AP-AP}$. In each deployment, STA positions are randomly placed. Traffic load is also adjusted to meet the abovementioned 90\% condition.

Fig.~\ref{Fig:randomDeployments} presents the $99^{\text{th}}$ and $50^{\text{th}}$ percentile delay for DCF, MAX2, and UNC under Poisson and Bursty traffic models and different inter-AP distances: in Fig.~\ref{Fig:randomDeployments}(a), $d_\text{AP-AP}=~10$ meters, while in Fig.~\ref{Fig:randomDeployments}(b), we double the distance to $d_\text{AP-AP}=~20$ meters. A consistent trend is observed in both cases: DCF consistently exhibits the highest delay, especially for Poisson traffic.

Furthermore, in Fig.~\ref{Fig:randomDeployments}(a), for Poisson traffic, MAX2 reduces the $99^{\text{th}}$ percentile delay by approximately 50\% (69.16 ms) with respect to DCF (137.66 ms), while UNC further improves performance, reducing the delay by 60\% compared to DCF. This performance gap narrows for Bursty traffic, where the difference between the MAPC approaches and DCF is approximately 31\%. Regarding the MAPC mechanisms, the distinction between them is primarily observed for Poisson traffic, with UNC exhibiting superior performance, achieving a 20\% improvement in the $99^{\text{th}}$ percentile delay. Their $50^{\text{th}}$ percentile delays are comparable and significantly lower than DCF's, demonstrating reductions of 52\% and 19\% for Poisson and Bursty traffic, respectively.

Finally, Fig.~\ref{Fig:randomDeployments}(b) examines a scenario where the inter-AP distance is doubled, creating more opportunities for large Co-SR groups. For Poisson traffic, UNC maintains the $99^{\text{th}}$ percentile packet delay below 6.24 ms, compared to 12.23 ms for MAX2 and 127.50 ms for DCF. Finally, for Bursty traffic, MAX2 reduces the $99^{\text{th}}$ percentile delay by 60\% relative to DCF, while UNC achieves a greater reduction of 72\%.

\section{Conclusions}\label{section:conclusions}

In this paper, we introduced a multi-AP coordination mechanism based on the formation of spatial reuse groups of AP-STA pairs, i.e., groups of devices that can transmit simultaneously. Our study compared the network delay of Co-SR against DCF. We focused on the dependency of the Co-SR on the devices' location (accounting for diverse inter-BSS spatial reuse interactions) and the performance when the group size is constrained according to TGbn's discussions. Our results highlighted the effectiveness of Co-SR group formation and its potential for enhancing overall network performance, demonstrating a significant delay reduction by the proposed Co-SR approaches over DCF, which ranges from 31\% to 95\%.

Our analysis invites for a future evaluation of different group creation algorithms, adaptive power control, and TXOP sharing strategies. It also encourages studying the effect of tuning key 802.11 parameters such as the set of allowed transmission rates, channel contention settings, and setting different transmission durations (e.g., per group) to add further degrees of freedom to the scheduling process.

\section{Acknowledgments}

D. Nunez, F. Wilhelmi and B. Bellalta were supported by grant Wi-XR PID2021-123995NB-I00 (MCIU/AEI/FEDER,UE), MdM CEX2021-001195-M (MICIU/AEI/10.13039/501100011033), and by SGR 00955-2021 AGAUR. 
G. Geraci was in part supported by the Spanish Research Agency through grants PID2021-123999OB-I00, CEX2021-001195-M, and CNS2023-145384. 
L. Galati Giordano was in part supported by UNITY-6G project, funded from European Union’s Horizon Europe Smart Networks and Services Joint Undertaking (SNS JU) research and innovation programme under the Grant Agreement No 101192650. 

\bibliographystyle{IEEEtran}
\bibliography{main}

\begin{thebibliography}{10}
\providecommand{\url}[1]{#1}
\csname url@samestyle\endcsname
\providecommand{\newblock}{\relax}
\providecommand{\bibinfo}[2]{#2}
\providecommand{\BIBentrySTDinterwordspacing}{\spaceskip=0pt\relax}
\providecommand{\BIBentryALTinterwordstretchfactor}{4}
\providecommand{\BIBentryALTinterwordspacing}{\spaceskip=\fontdimen2\font plus
\BIBentryALTinterwordstretchfactor\fontdimen3\font minus \fontdimen4\font\relax}
\providecommand{\BIBforeignlanguage}[2]{{%
\expandafter\ifx\csname l@#1\endcsname\relax
\typeout{** WARNING: IEEEtran.bst: No hyphenation pattern has been}%
\typeout{** loaded for the language `#1'. Using the pattern for}%
\typeout{** the default language instead.}%
\else
\language=\csname l@#1\endcsname
\fi
#2}}
\providecommand{\BIBdecl}{\relax}
\BIBdecl

\bibitem{WBA_2025_report}
{Wireless Broadband Alliance}, ``{WBA} {Annual Industry Report} 2025,'' \url{https://wballiance.com/resource/wba-annual-industry-report-2025/}, December 2024, accessed on 28/02/2025.

\bibitem{OugGerPol2024}
E.~Oughton, G.~Geraci, M.~Polese, M.~Ghosh, W.~Webb, and D.~Bubley, ``The future of wireless broadband in the peak smartphone era: {6G}, {Wi-Fi} 7, and {Wi-Fi} 8,'' \emph{techrxiv.173221414.41000458}, 2024.

\bibitem{UHRobjectives}
``{IEEE 802.11-22/0078r3: 802.11 UHR Draft Proposed PAR},'' \url{https://mentor.ieee.org/802.11/dcn/23/11-23-0078-03-0uhr-uhr-draft-proposed-par.docx}, January 2023, accessed on January 2024.

\bibitem{AdrianGarciaSurvey}
A.~Garcia-Rodriguez, D.~López-Pérez, L.~Galati-Giordano, and G.~Geraci, ``{IEEE 802.11be: Wi-Fi 7 Strikes Back},'' \emph{IEEE Communications Magazine}, vol.~59, no.~4, pp. 102--108, 2021.

\bibitem{UPF_Nokia_wifi8}
L.~Galati-Giordano, G.~Geraci, M.~Carrascosa, and B.~Bellalta, ``{What will Wi-Fi 8 Be? A Primer on IEEE 802.11bn Ultra High Reliability},'' \emph{IEEE Communications Magazine}, vol.~62, no.~8, pp. 126--132, 2024.

\bibitem{MAPC_mentor_MediaTek_1217r2}
K.~Lu, J.~Yee, F.~Hsu, L.-H. Sun, Y.~Fang, and G.~Bajko, ``{Multi-AP Coordinaton Setup Scheme},'' Aug 2024, doc.: IEEE 802.11-24/1217r2.

\bibitem{MAPCframework_mentor_LG_1514r1}
G.~Kim, I.~Jang, J.~Choi, S.~Baek, Y.~Yoon, D.~Cha, H.~Lee, E.~Park, D.~Lim, J.~Chun, I.~Jung, H.~Cho, and S.~Kim, ``{Multi-AP Framework for C-SR},'' Nov 2024, doc.: IEEE 802.11-24/1514r1.

\bibitem{Val_2025}
\BIBentryALTinterwordspacing
I.~Val, D.~López-Pérez, A.~Kijanka, S.~Schelstraete, L.~Muñoz, D.~Arlandis, and M.~Martínez, ``{Wi-Fi 8 Unveiled: Key Features, Multi-AP Coordination, and the Role of C-TDMA},'' Mar. 2025. [Online]. Available: \url{http://dx.doi.org/10.36227/techrxiv.174114571.17876683/v1}
\BIBentrySTDinterwordspacing

\bibitem{haxhibeqiri2024coordinated}
J.~Haxhibeqiri, X.~Jiao, X.~Shen, C.~Pan, X.~Jiang, J.~Hoebeke, and I.~Moerman, ``{Coordinated SR and restricted TWT for time sensitive applications in WiFi 7 networks},'' \emph{IEEE Communications Magazine}, vol.~62, no.~8, pp. 118--124, 2024.

\bibitem{mypaper_throughputAnalisis}
D.~Nunez, F.~Wilhelmi, L.~Galati-Giordano, G.~Geraci, and B.~Bellalta, ``{Spatial Reuse in IEEE 802.11bn Coordinated Multi-AP WLANs: A Throughput Analysis},'' in \emph{{2024 IEEE Conference on Standards for Communications and Networking (CSCN)}}, 2024, pp. 265--270.

\bibitem{mio_MAPC_groups_scheduling}
D.~Nunez, M.~Smith, and B.~Bellalta, ``{Multi-AP Coordinated Spatial Reuse for Wi-Fi 8: Group Creation and Scheduling},'' in \emph{2023 21st Mediterranean Communication and Computer Networking Conference (MedComNet)}, 2023, pp. 203--208.

\bibitem{TXOPsharingPaper}
D.~Nunez, F.~Wilhelmi, S.~Avallone, M.~Smith, and B.~Bellalta, ``{TXOP} sharing with coordinated spatial reuse in multi-{AP} cooperative {IEEE} 802.11be {WLANs},'' in \emph{2022 IEEE 19th Annual Consumer Communications Networking Conference (CCNC)}, 2022, pp. 864--870.

\bibitem{francesc_wifi8_markov}
F.~Wilhelmi, L.~Galati-Giordano, G.~Geraci, B.~Bellalta, G.~Fontanesi, and D.~Nuñez, ``{Throughput Analysis of IEEE 802.11bn Coordinated Spatial Reuse},'' in \emph{2023 IEEE Conference on Standards for Communications and Networking (CSCN)}, 2023, pp. 401--407.

\bibitem{pathloss}
S.~Merlin, G.~Barriac, H.~Sampath, L.~Cariou, T.~Derham, J.-P.~L. Rouzic, R.~Stacey, M.~Park, C.~Ghosh, R.~Porat, N.~Jindal, Y.~Inoue, Y.~Asai, Y.~Takatori, A.~Kishida, and A.~Yamada, ``{TGax Simulation Scenarios},'' Nov. 2015, doc.: IEEE 802.11-14/0980r16.

\bibitem{lee2007CaptureEffect}
J.~Lee, W.~Kim, S.-J. Lee, D.~Jo, J.~Ryu, T.~Kwon, and Y.~Choi, ``An experimental study on the capture effect in 802.11a networks,'' in \emph{Proc. ACM WiNTECH}, 2007, pp. 19--26.

\end{thebibliography}

\end{document}